# Directional infrared emission resulting from cascade population inversion and four-wave mixing in Rb vapours

Alexander Akulshin,[1,*] Dmitry Budker,[2] and Russell McLean[1]

[1]*Centre for Atom Optics and Ultrafast Spectroscopy,
Swinburne University of Technology, PO Box 218, Hawthorn 3122, Australia*
[2]*Department of Physics, University of California, Berkeley, CA 94720-7300, USA*

*Corresponding author:  aakoulchine@swin.edu.au*

Directional infrared emission at 1.37 and 5.23 µm is generated in Rb vapours that are step-wise excited by low-power resonant light. The mid-infrared radiation originating from amplified spontaneous emission on the $5D_{5/2} \to 6P_{3/2}$ transition consists of forward- and backward-directed components with distinctive spectral and spatial properties. Diffraction limited near-infrared light at 1.37 µm generated in the co-propagating direction only is a product of parametric wave mixing around the $5P_{3/2} \to 5D_{5/2} \to 6P_{3/2} \to 6S_{1/2} \to 5P_{3/2}$ transition loop. This highly non-degenerate mixing process involves one externally applied and two internally generated optical fields. Similarities between wave mixing generated blue and near-IR light are demonstrated.

Parametric four-wave mixing (FWM) is at the heart of many wavelength conversion processes. Although it was originally discovered using high-power pulsed lasers, under suitable conditions parametric FWM can also occur with low-power cw lasers, making very high spectral resolution of new spectroscopic features possible

Wave mixing of low-intensity resonant radiation enhanced by light-induced ground-state coherence in alkali vapours can generate new mutually coherent optical fields having comb-like spectra with spacing in the radio frequency [1] or microwave [2] spectral range. In the present work we investigate a situation where optical coherence results in new coherent-field generation with frequency difference in optical domain.

Frequency up-conversion of low-power cw laser radiation into highly coherent and directional blue light in Rb and Cs vapours is an active area of research with numerous potentially important applications [3,4,5,6,7,8]. Stepwise excitation of transitions in alkali vapours not only provides conditions for amplified spontaneous emission, but also creates strong cavity-free spatial anisotropy for new field generation through the phase matching condition.

Alkali atoms driven by bi-chromatic co-propagating laser radiation tuned close to step-wise transitions, $5S_{1/2} \to 5P_{3/2}$ and $5P_{3/2} \to 5D_{5/2}$ in the case of Rb atoms (Fig. 1), can produce population inversion on the mid-IR $5D_{5/2} \to 6P_{3/2}$ transition, since the spontaneous lifetime of the $5D_{5/2}$ level is 238.5 ns compared to 120.7 ns for the $6P_{3/2}$ level [9,10]. The 5.23 µm radiation, which originates from amplified spontaneous emission, can be randomly directed in a dense atomic sample [11]; however, in the case of a pencil-shaped interaction region, sufficiently high excitation rate, and modest atomic density ($5 \times 10^{11}$ cm$^{-3}$ < $N$ < $1 \times 10^{13}$ cm$^{-3}$), the radiation consists of collimated forward- and backward-propagating components. Mixing of the forward-directed mid-IR component with the applied laser fields produces collimated blue light (CBL) in the co-propagating direction, as only in this case is the phase-matching relation satisfied. Recently it was shown that CBL can also be generated with cw single-frequency two-photon excitation [12,13].

Atoms on the $6P_{3/2}$ level can be involved in the processes of blue light generation and spontaneous decay to the $6S_{1/2}$ level. The branching ratio for this 2.73 µm decay channel is 0.49 compared to 0.31 for the blue transition [14].

Because the lifetime of the $6P_{3/2}$ level is more than twice that of the $6S_{1/2}$ level, a substantial population inversion is possible on the $6P_{3/2} \to 6S_{1/2}$ transition, leading to directional IR light generation at 2.73 µm due to amplified spontaneous emission.

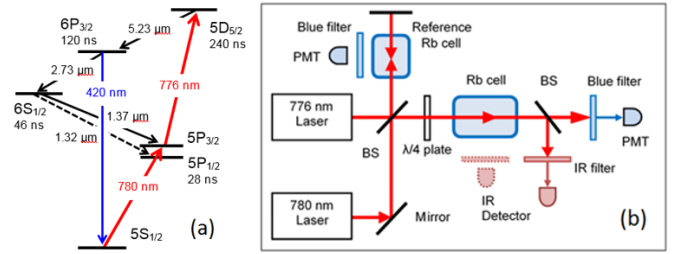

Fig. 1. (a) Rb atom energy levels involved in frequency up- and down-conversion. (b) Schematic diagram of the experimental set-up.

The combination of the laser light at 776 nm and the two internally generated IR components at 5.23 and 2.73 µm may produce optical coherence between the $6S_{1/2}$ and $5P_{3/2}$ levels, and establish another loop for parametric FWM. In earlier papers collimated backward mid-IR radiation at 5.23 µm was not detected [3], while co-propagating near-IR emission at 1.37 µm was mentioned only briefly [15]. Here we undertake an experimental study of the spectral and spatial properties of both the

mid- and near-IR radiation generated in a specially fabricated Rb vapour cell.

The basic optical scheme of the experiment is shown in Fig. 1b. Extended cavity diode laser (ECDL) sources of coherent light at 780 and 776 nm are used for step-wise excitation of the Rb atoms. The optical frequency of the 780 nm ECDL is either scanned across the $^{85}$Rb D$_2$ absorption line or modulation-free stabilized to the Doppler-free polarization-spectroscopy resonance of the $5S_{1/2}(F=3) \rightarrow 5P_{3/2}(F'=4)$ transition obtained in an auxiliary Rb cell. The 776 nm laser is also either scanned across the $^{85}$Rb $5P_{3/2} \rightarrow 5D_{5/2}$ transition or modulation-free locked to a low-finesse tuneable reference Fabry-Perot cavity. The power and polarizations of the 780 nm and 776 nm laser light are controlled with wave plates and polarizers. Radiation from the two ECDLs is combined on a beam splitter to form two bi-chromatic beams. Doppler-free blue fluorescence resonances obtained with counter-propagating bi-chromatic radiation in a reference vapor cell provide convenient fine frequency references [16].

The temperature of the 5 cm-long specially fabricated Rb vapor cell with sapphire windows, containing a natural mixture of Rb isotopes and no buffer gas, is set within the range 50-100°C, so that the atomic density $N$ of saturated Rb vapor varies between $1 \times 10^{11}$ cm$^{-3}$ and $6 \times 10^{12}$ cm$^{-3}$. Sapphire is transparent for near-IR and partially transparent for mid-IR radiation at 5.23 μm. The maximum powers of the 780 and 776 nm light beams in the cell are approximately 15 and 5 mW, respectively. The bi-chromatic beam can be focused into the vapour cell with a long focal length lens to reduce the cross section of the interaction region to a minimum of $\approx 1$ mm$^2$.

Interference filters of optical density approximately 0.5 and 4.0 at 420 nm and 780 nm respectively, are used to spectrally separate the CBL from the laser light. We detect the blue radiation using photomultiplier tubes (PMTs) and a CCD camera. Near- and mid-IR radiation is detected with a Ge photodiode and a Ge:Au photoresistor cooled with liquid nitrogen, respectively, while laser light is blocked by band-pass filters. The signal-to-noise ratio of the signals from the IR detectors is enhanced by 500 Hz mechanical chopping of the applied laser light at 776 nm and subsequent lock-in amplification. The signals proportional to blue fluorescence or collimated blue light, with the exception of Fig. 2a, are recorded without lock-in processing. The directly recorded signals consist of pulses, which are unresolved in the case of slow frequency scanning. For this reason the corresponding dependences appear as solid profiles (Fig. 2b and Fig. 4).

Collimated forward-directed mid-IR (FMIR) radiation generated in Rb vapours excited by a bi-chromatic laser beam was detected for the wide range of experimental parameters for which the CBL is also present. We find that maxima of the FMIR emission and isotropic blue fluorescence, which is proportional to the number of excited atoms, occur at the same applied laser frequencies, however, the FMIR occurs over a smaller (by about a factor of two) range of the frequency detuning of the 776 nm laser.

Typical dependences of collimated 5.23 μm and blue radiation intensities as a function of the second-step laser frequency in the vicinity of the $5P_{3/2} \rightarrow 5D_{5/2}$ transition, while the fixed-frequency 780 nm laser is locked to the $5S_{1/2}(F=3) \rightarrow 5P_{3/2}(F'=4)$ transition, are shown in Fig. 2a. The difference in the two profiles reflects their distinctive (parametric and non-parametric) origins. The divergence of the FMIR emission, as well as the CBL, is approximately 6 mrad and diffraction limited.

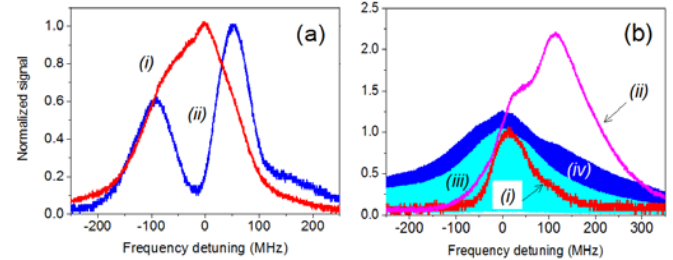

Fig. 2. (a) Intensity of forward-propagating mid-IR radiation and CBL as a function of the 776 nm laser frequency detuning, while the 780 nm laser is locked to the $5S_{1/2}(F=3) \rightarrow 5P_{3/2}(F'=4)$ transition. Atomic density $N$ in the cell is approximately $2.5 \times 10^{12}$ cm$^{-3}$. (b) Backward mid-IR radiation (profiles *i* and *ii*) and blue fluorescence (*iii* and *iv*) at different first-step laser power as a function of the 776 nm laser frequency. The power at 776 nm is 1.6 mW, Profiles *i* and *iii* are taken with 2.9 mW at 780 nm, while for *ii* and *iv* profiles the power at 780 nm is 5 mW.

Directional mid-IR emission has been also detected in the backward direction. We find that this component of the collimated 5.23 μm radiation reveals strong dependence on the applied power at 780 nm as shown in Fig. 2b. The spectral profile of the backward-directed mid-IR (BMIR) emission shifts to higher frequencies at higher laser power.

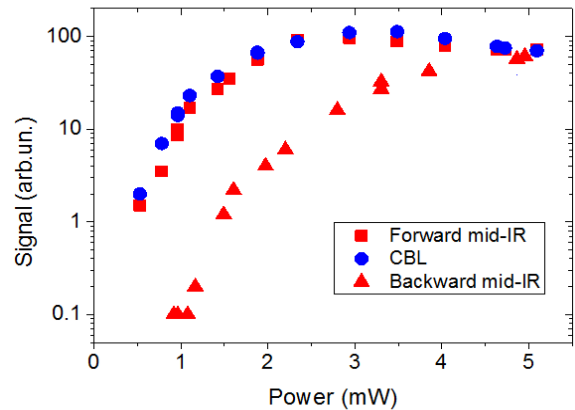

Fig. 3. Collimated blue, forward- and backward-directed mid-IR radiation generated in Rb vapours at different power of the first-step laser at 780 nm, while the atomic density in the cell is approximately $3 \times 10^{12}$ cm$^{-3}$.

Figure 3 demonstrates that the intensity dependences of the CBL and FMIR emission on the applied laser power are similar, while the intensity variation of the BMIR emission is quite different. At relatively low applied laser power ($P < 1.5$ mW) there is large imbalance between the intensity of backward- and forward-directed emission.

However, the BMIR grows faster with the applied laser power at 780 nm matching with the FMIR at 6 mW. It is likely that different intensities in the co- and counter-propagating directions at low laser power is the reason why the collimated backward radiation at 5.23 μm was not observed in [13].

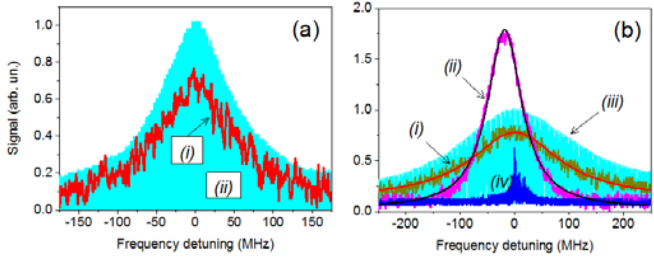

Fig. 4. (a) Near-IR emission detected with the on-axis photodiode *i* and anisotropic blue fluorescence *ii* for atomic density $N \approx 7\times10^{11}$ cm$^{-3}$ as a function of the 776 nm laser detuning from the maximum Rb excitation to the $5D_{5/2}$ level, while the 780 nm laser is locked to the $5S_{1/2}(F=3)\rightarrow 5P_{3/2}(F'=4)$ transition. Laser light is not focused. (b) Curves *i* and *ii* show near-IR emission detected with the off-axis and on-axis photodiodes, respectively, for $N \approx 7\times10^{12}$ cm$^{-3}$. Curves *iii* and *iv* demonstrate blue fluorescence detected in the main Rb cell and in the reference cell, respectively. Curves *i* and *ii* are fitted to Lorentzian function.

For certain experimental conditions, such as the applied laser light being unfocused in the Rb cell and modest atomic density ($N < 10^{12}$ cm$^{-3}$), the CBL is well established but the gain on the $6P_{3/2}\rightarrow 6S_{1/2}$ transition due to population inversion is not high enough to produce intense and directional radiation at 2.73 μm. In this case atoms from the $6S_{1/2}$ level decay spontaneously to the $5P_{1/2}$ and $5P_{3/2}$ levels with branching ratio 0.34 and 0.66, respectively, emitting photons at 1.32 and 1.37 μm. Figure 4a shows the near-IR emission in the co-propagating direction and blue fluorescence detected simultaneously in the transverse direction. The dependences of the near-IR emission and blue fluorescence on laser detuning are similar, suggesting that spontaneous decay is the common origin of both radiations. In this case both the near-IR components at 1.32 and 1.37 μm contribute to the observed signal. As population inversion on the $6S_{1/2}\rightarrow 5P_{1/2}$ transition is achieved much more easily than on the $6S_{1/2}\rightarrow 5P_{3/2}$ transition, we expect that the emission at 1.32 μm in the co- and counter-propagating directions is comparable to or even stronger than the 1.37 μm emission, despite the branching ratios. The full width at half maximum of the profiles is less than the Doppler width because of velocity selectivity of the step-wise excitation.

The situation is different above the threshold for collimation (and intensity) of the 2.73 μm radiation. Then the combined action of the laser light at 776 nm and the two co-propagating internally generated IR fields at 5.23 μm and 2.73 μm is able to produce coherence on the 1.37 μm $6S_{1/2}\rightarrow 5P_{3/2}$ transition, establishing a new pathway for parametric FWM. As a result, the emission on this transition is enhanced and becomes highly anisotropic, as the phase-matching relation $\boldsymbol{k}_{NIR} = \boldsymbol{k}_2 + \boldsymbol{k}_3 - \boldsymbol{k}_4$ can be satisfied for certain directions only, where $\boldsymbol{k}_{2,3,4}$ are the wave vectors of the fields at 776 nm, 5.23 μm and 2.73 μm, respectively, and $\boldsymbol{k}_{NIR}$ is the wave vector of the emission at 1.37 μm.

Typical powers of the forward-directed near-IR emission detected by slightly off-axis and distant on-axis photodiodes as a function of the 776 nm laser detuning are shown in Fig. 4b. Following [16], we use Doppler-free blue fluorescence resonances obtained in the reference cell as convenient absolute frequency references. The largest peak on the two-photon $5S_{1/2}(F=3)\rightarrow 5D_{5/2}(F''=5)$ transition, curve (*iv*), spectrally coincides with the maximum of the broader blue fluorescence profile in the main cell when the 780 nm laser is locked to the $5S_{1/2}(F=3)\rightarrow 5P_{3/2}(F'=4)$ transition. The enhanced amplitude, narrowing and frequency shift of the profile that represents the co-propagating radiation relative to the off-axis radiation indicate the onset of the new coherent and directional radiation. We also find that spectral dependences of the slightly off-axis, as well as anisotropic near-IR emission and blue fluorescence are similar, reflecting their common 1predominantly incoherent origin. Fitting Lorentzian functions to the profiles shows that the width of the more divergent incoherent component is 200 MHz compared to 76 MHz for the collimated forward-directed near-IR (FNIR) emission. The red frequency shift of the FNIR profile, under the present experimental conditions, is approximately 20 MHz.

The forward-directed near-IR emission consists of amplified spontaneous emission at 1.32 μm and the FWM component at 1.37 μm with different intensity and spatial properties as has been shown using an optical spectrum analyzer (ANDO AQ-6315E). The more collimated component at 1.37 μm is more efficiently coupled into an optical fibre and is unambiguously detected (Fig. 5a), while the radiation at 1.32 μm is below the detection threshold.

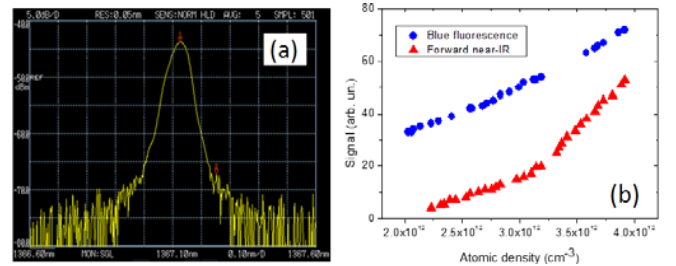

Fig. 5. (a) Spectrum of the collimated forward-directed near-IR emission. (b) Atomic density dependences of the FNIR radiation and blue fluorescence.

The FNIR emission and CBL share several common characteristic features attributable to their common origin, including threshold dependence on atomic density and laser power, and sensitivity to the polarization of the applied laser light. Both near-IR and blue light are highly directional and the direction is determined by the phase-matching condition. The divergences of the FNIR radiation and the CBL are similar (approximately 6 mrad) and are both controlled by the applied laser light. Backward-directed emission in the near-IR and blue spectral regions is much weaker and more divergent.

The atom density has a strong effect on the intensity of the FNIR radiation. Figure 5b demonstrates that both the FNIR radiation and blue fluorescence grow steadily with atom density $N$. However, under the present experimental conditions, the FNIR radiation grows faster above $N \approx 3.3\times10^{12}$ cm$^{-3}$. This is again consistent with the transition from spontaneously produced emission to generation, via a wave-mixing process, of collimated light in the forward direction.

The intensity of the FNIR radiation depends on the intensity of the applied laser light. It grows steadily with the 776 nm power. At the same experimental conditions ($N \approx 3.7\times10^{12}$ cm$^{-3}$) the measured power of FNIR and CBL is approximately 70 and 400 nW, respectively. Comparing the FNIR and blue fluorescence variations with the applied laser power we find that the collimated near-IR emission grows faster than the number of atoms excited to the $5D_{5/2}$ level. This is expected behaviour given that two optical fields involved in the parametric FWM process and FNIR generation originate from amplified spontaneous emission and should possess threshold-like power dependences.

In conclusion, we experimentally studied the spectral and spatial properties of the mid- and near-IR emission generated in Rb vapours simultaneously with coherent blue light. Collimated mid-IR radiation at 5.23 µm was detected in the co- and counter-propagating directions, as expected from the ASE mechanism. However, the distinctive spectral and spatial characteristics of the forward and backward-directed mid-IR radiation suggest that beside amplified spontaneous emission, which is possible due to favorable lifetime ratios for the relevant states, additional nonlinear processes contribute to the mid-IR generation. Understanding the role of various parametric and nonparametric processes, which are responsible for efficiency of the backward-directed emission in dilute atomic samples, should help in assessing the usefulness of this approach for remote sensing and laser-guide-star techniques [17,18] that exploit efficient and collimated backward emission. Another intriguing question that might be explored is the common origin of forward directed mid-IR radiation at 5.23 and 2.73 µm.

Using a fairly standard arrangement to generate CBL in alkali vapours, we have demonstrated that a parametric four-wave mixing process can also occur in Rb vapours on the $5P_{3/2} \rightarrow 5D_{5/2} \rightarrow 6P_{3/2} \rightarrow 6S_{1/2} \rightarrow 5P_{3/2}$ loop. Remarkably, only one transition, the first in the loop, is driven by laser radiation, while the other fields are self-generated in Rb vapours. We reveal a number of similarities between the spectral and propagation characteristics of the 1.37 µm radiation produced in this process and the blue light, which support the wave-mixing origin.

The similarities also suggest that the linewidths of the near-IR and mid-IR emissions may be quite small as was shown for the collimated blue light [19]. This, together with the fact that the approach followed here is applicable to a wide range of atomic media, would thus represent a new way of generating coherent and frequency tuneable light in the important mid-IR spectral region, meriting further investigation.

We thank B. Patton for making the Rb cell with sapphire windows, A. Kireev for supplying a mid-IR detector and I. Novikova and E. Mikhailov for useful discussions. DB is grateful to the Centre for Atom Opticas and Ultrafast Spectroscopy at the Swinburne University of Technology for hosting him as a Distinguished Visiting Researcher.

1. A. M. Akulshin, A Cimmino, A. I. Sidorov, R. McLean, and P. Hannaford, *J. Opt. B: Quantum Semiclass. Opt.* **5,** S479 (2003).
2. M. M. Kash, V. A. Sautenkov, A. S. Zibrov, L. Hollberg, G. R. Welch, M. D. Lukin, Y. Rostovtsev, E. S. Fry, and M. O. Scully, *Phys. Rev. Lett.* **82**, 5229 (1999).
3. T. Meijer, J. D. White, B. Smeets, M. Jeppesen, and R. E. Scholten, Opt. Lett. **31**, 1002 (2006).
4. A. S. Zibrov, M. D. Lukin, L. Hollberg, and M. O. Scully, *Phys. Rev. A* **65**(5), 051801 (2002).
5. A. M. Akulshin, R. J. McLean, A. I. Sidorov, and P. Hannaford, *Optics Express* **17**(25), 22861 (2009).
6. A. Vernier, S. Franke-Arnold, E. Riis, and A. S. Arnold, *Optics Express* **18**(16), 17020 (2010).
7. J. T. Schultz, S. Abend, D. Döring, J. E. Debs, P. A. Altin, J. D. White, N. P. Robins, and J. D. Close, *Opt. Lett.* **34**, 2321 (2009).
8. G. Walker, A.S. Arnold, S. Franke-Arnold, *Phys. Rev. Lett.* **108**, 243601 (2012).
9. D. Sheng, A. Perez Galvan, and L. A. Orozco, *Phys. Rev. A* **78**, 062506 (2008).
10. E. Gomez, S. Aubin, L. A. Orozco, and G. D. Sprouse, *J. Opt. Soc. Am.* B **21**, 2058–2067 (2004).
11. A. I. Lvovsky, S. R. Hartmann and F. Moshary, *Phys. Rev. Lett.* **82**, 4420 (1999).
12. C. V. Sulham, G. A. Pitz, and G. P. Perram, *Appl. Phys.* B **101,** 57 (2010).
13. E. Brekke and L. Alderson, *Opt. Lett.* **38**, 2147 (2013).
14. O. S. Heavens, *J. Opt. Soc. Am.* **51**(10), 1058–1061 (1961).
15. A. S. Zibrov, L. Hollberg, V. L. Velichansky, M. O. Scully, M. D. Lukin, H. G. Robinson, A. B. Matsko, A. V. Taichenachev, and V. I. Yudin*, CP551 Atomic Physics* **17** ed. By E. Arimondo, P. DeNatale, and M. Inguscio, 204 (2001).
16. A. M. Akulshin, B. V. Hall, V. Ivannikov, A. A. Orel, and A. I. Sidorov, *J. Phys. B: At. Mol. Opt. Phys.* **44,** 215401 (2011).
17. W. Happer, G. J, MacDonald, C. E. Max, F. J. J. Dyson, *Opt. Soc. Am.* A **11**, 263 (1994).
18. A. Akulshin, D. Budker, B. Patton, R. McLean, arXiv:1310.2694 [physics.atom-ph], (2013).
19. A. Akulshin, Ch. Perrella, G-W. Truong, R. McLean, and A. Luiten*, J. Phys. B* **45**, 245503 (2012).